\documentclass[aps,prb,twocolumn,citeautoscript,nofootinbib,10pt]{revtex4-1}  
\pdfoutput=1

\usepackage{amsmath}
\usepackage{graphicx}
\usepackage{color}
\usepackage{amssymb}

\usepackage[tight]{subfigure} 

\usepackage{color} 
\usepackage[papersize={8.5in,11in}]{geometry}
\usepackage[colorlinks=true]{hyperref}
\hypersetup{
    bookmarks=true,         
    unicode=false,          
    pdftoolbar=true,        
    pdfmenubar=true,        
    pdffitwindow=false,     
    pdfstartview={FitH},    
    pdftitle={RMI},    
    pdfauthor={Author},     
    pdfsubject={Subject},   
    pdfcreator={Creator},   
    pdfproducer={Producer}, 
    pdfkeywords={keyword1} {key2} {key3}, 
    pdfnewwindow=true,      
    colorlinks=true,       
    linkcolor=magenta, 
    citecolor=blue,        
    filecolor=magenta,      
    urlcolor=blue           
} 

\newcommand{\beq}{\begin{equation}}
\newcommand{\eeq}{\end{equation}}
\newcommand{\ben}{\begin{enumerate}}
\newcommand{\een}{\end{enumerate}}

\def\be{\begin{eqnarray}}
\def\ee{\end{eqnarray}}

\begin{document}

\title{Critical scaling of the mutual information in two-dimensional disordered Ising models}

\author{P. V. Sriluckshmy}
\affiliation{Max-Planck Institute for the Physics of Complex Systems\\
Noethnitzer Str.  38, 01187, Dresden, Germany}

\author{Ipsita Mandal}
\affiliation{Max-Planck Institute for the Physics of Complex Systems\\
Noethnitzer Str.  38, 01187, Dresden, Germany}

\date{\today}

\begin{abstract}
R\'enyi Mutual information (RMI), computed from second R\'enyi entropies, can identify classical phase transitions from their finite-size scaling at the critical points. We apply this technique to examine the presence or absence of finite temperature phase transitions in various two-dimensional models on a square lattice, which are extensions of the conventional Ising model by adding a quenched disorder. When the quenched disorder causes the nearest neighbor bonds to be both ferromagnetic and antiferromagnetic, (a) a spin glass phase exists only at zero temperature, and (b) a ferromagnetic phase exists at a finite temperature when the antiferromagnetic bond distributions are sufficiently dilute. Furthermore, finite temperature paramagnetic-ferromagnetic transitions can also occur when the disordered bonds involve only ferromagnetic couplings of random strengths. In our numerical simulations, the ``zero temperature only" phase transitions are identified when there is no consistent finite-size scaling of the RMI curves, while for finite temperature critical points, the curves can identify the critical temperature $T_c$ by their crossings at $T_c$ and $2\, T_c $.
\end{abstract}

\maketitle

\section{Introduction}


A system with quenched disorder is one where some of the parameters defining its behaviour are random variables that do not evolve with time, and as such they are quenched or frozen. One such example is the spin glass, which is a disordered magnetic system. Although the interactions between the magnetic moments in a spin glass are frustrated, nevertheless it exhibits a transition from the paramagnetic phase to a novel ordered phase -- the spin glass phase -- where the the spins are “frozen” in an irregular pattern. Since randomness and frustration are the necessary ingredients for the emergence of a spin glass, quenched disorder in a classical spin system can lead to frustration and thus generate spin glasses. In two-dimensional (2d) lattices, there have been extensive studies of such systems and all the results point to the fact that no spin glass phase exists for nonzero temperatures \cite{young-review}. In fact, the lower critical dimension for Ising spin glasses is believed to be two, which indicates that the critical temperature for 2d is zero.
However, finite temperature paramagnetic-ferromagnetic transitions are possible in such disordered systems if the frustration is reduced \cite{nishimori,matthew-fisher}, or removed altogether \cite{fisch,jayaprakash,zobin,levy}. This can be achieved either by sufficiently reducing the number of antiferromagnetic bonds, or having only ferromagnetic couplings in the model.

In this work, we study a variety of classical spin-$1/2$ models on square lattices, which are extensions of the Ising model with some kind of quenched disorder. We reexamine the presence or absence of finite temperature phase transitions for these models using the method of classical Monte Carlo simulations, which, via a replica-trick scheme, can detect finite temperature critical points, even identifying their universality classes without any {\it a priori} knowledge of an order parameter or associated broken symmetry \cite{melko2010,Singh,WL}. The method involves computation of R\'enyi mutual information (RMI) derived from the second R\'enyi entropies. The critical scaling of this mutual information with system size can detect and classify phase transitions. This method has been successfully applied in a number of classical systems \cite{stephen2013,stephan2014,troyer,vidal,Alba1,Alba2,stephen2016,Johannes,ipsita-roger}. The physical reason for information quantities to be able to detect phase transitions is the deep connection between certain measurable thermodynamic quantities
and principles of information theory. In fact, information can be quantified in terms of entropy, which in turn can be defined from thermodynamic observables \cite{shannon,cardy}. 
For classical phase transitions, correlation lengths diverge at the critical points, indicative of the existence of long-range channels for information transfer. Furthermore, the usefulness of the mutual information was demonstrated in a striking way by St\'ephan {\it et. al.} \cite{stephan2014}, where simple classical Monte Carlo simulations of the 2d Ising model at its phase transition
could compute the central charge of the associated $(1+1)$-dimensional conformal field theory (CFT) \cite{belavin,friedan,wilczek,kitaev,cardy}, thus identifying its universality class.

In the previous studies \cite{stephen2013,stephan2014,troyer,vidal,Alba1,Alba2,stephen2016,Johannes,ipsita-roger}, no disorder was involved. Hence, it is not clear a priori whether the results derived there would hold for disordered systems, given the fact that there are considerable error bars for disorder realizations. The analytical expressions
and scaling ansatz used for the clean cases need to be rederived in the presence of disorder.
However, our numerical simulations indicate that the analytical arguments should also hold
in presence of disorder.

The paper is structured as follows: In Sec.~\ref{model}, we discuss the models for which the RMI is calculated as a function of system size and temperature. In Sec.~\ref{method}, we describe the numerical techniques employed to get the RMI curves. Sec.~\ref{results} describe our numerical results. We conclude with some discussion and outlook in Sec.~\ref{conclude}.


\section{Models}
\label{model}

In this section, we describe the 2d classical spin models on a square lattice of linear dimension $L$, for which we probe the critical temperature $T_c$. Throughout this work, we will set the energy scale $J= 1$.

The Edwards-Anderson spin glass model \cite{edwards}, described by the Hamiltonian:
\begin{align} 
H = - \sum_{\langle i j \rangle} J_{ij} \,S^z_i\, S^z_j \, ,
\label{EA-model}
\end{align}
has a glassy phase below $T_c$.
Here $S^z_i = \pm 1 $ represents the $L^2$ Ising spins and $ \langle i j \rangle $ denotes nearest neighbor sites. The coupling strength $J_{ij} $  between the nearest neighbors is a random (quenched) variable. The values of $J_{ij}$ are drawn from a continuous Gaussian distribution with zero mean and unit variance:
\begin{align}
\label{prob1}
P(J_{ij}) = \frac{\exp \left(  -J_{ij}^2   / 2 \right) }{\sqrt{2\, \pi} } \,,
\end{align}
or from a discrete bi-modal distribution:
\begin{align}
P(J_{ij}) = \frac{1}{2} \left [ \delta (J_{ij} +1) + \delta (J_{ij} -1) \right   ] ,
\end{align}
such that $J_{ij} = \pm 1$ with equal probability. For 2d, all studies found $T_c$ to be zero.

When the values of $J_{ij}$ in Eq.~(\ref{EA-model}) are drawn from the probability distribution
\begin{align}
P(J_{ij}) = p \,\delta (J_{ij} +1) +(1-p) \,\delta (J_{ij} -1) \,, \text{ with } 0\leq p\leq 1\,,
\label{randomising}
\end{align}
then we get a random-bond Ising model, where the probabilities $p$ and $(1-p)$ are associated with antiferromagnetic and ferromagnetic couplings, respectively. It was shown in Ref.~\onlinecite{nishimori,matthew-fisher} that although the spin glass phase does not exist in 2d for $T>0$ ( at $T=0$, the model has a spin glass phase), but for a sufficiently small concentration of antiferromagnetic bonds ($p<p_c$), there exists a finite temperature phase transition from the paramagnetic to the ferromagnetic phase. At $p= 0$, we have the standard ferromagnetically coupled 2d
Ising model with $T_c \simeq 2.269$. For $p >0$, this $T_c$ is reduced by frustration induced by the  antiferromagnetic couplings, until it vanishes at $p_c \simeq 0.12$ \cite{aoki}.

We also consider a variation of the Hamiltonian in Eq.~(\ref{EA-model}) by adding a second nearest
neighbor interaction of uniform strength $ J'$, such that the Hamiltonian is given by:
\begin{align}
H = - \sum_{\langle i j \rangle} J_{ij} \,S^z_i\, S^z_j 
- J'\sum_{\langle\langle i j \rangle\rangle }   S^z_i \,S^z_j \,  .
\label{ferro-model}
\end{align}
Here  $\langle\langle i j \rangle\rangle $ denotes next nearest neighbor sites with distance $\sqrt{2}$ lattice spacing units apart. The system, with $J' = 1$ and $J_{ij} = \pm \lambda $ (dubbed as ``randomly coupled ferromagnet" or RCF), was predicted to have a finite temperature phase transition from numerical analysis \cite{ferro1,ferro2,ferro3}. The authors estimated the  ordering temperatures to be close to $2$ for $\lambda  = 0 . 5,\, 0.7$ and dropping to zero near $ \lambda = 1$. 

We also study a 2d random-bond Ising model only with ferromagnetic couplings, such that in the Hamiltonian in Eq.~(\ref{EA-model}), we now have \cite{fisch,jayaprakash,zobin,levy}:
\begin{align}
P(J_{ij}) = p\, \delta (J_{ij} -1) +(1-p) \,\delta (J_{ij}-\tilde J  )  \,, \text{ with } 0\leq p\leq 1\,. \label{diffcouplings}
\end{align}
For $\tilde J=0$,  the resulting ``dilute ferromagnet" is a disorder model where nonzero exchange interactions exist only between a fraction $p$ of neighboring pairs of Ising spins.
This system shows a finite temperate paramagnetic-ferromagnetic phase transition for $1/2<p \leq 1$.
The dependence of $T_c$ on $p$ can be found in  Fig.~2 of Ref.~\onlinecite{jayaprakash,zobin} and Table~2 of Ref.~\onlinecite{levy}.
Furthermore, for $p=1/2$, the model is self-dual and $T_c$ is determined exactly by the relation \cite{fisch,derrida}:
\begin{align}
\label{analytic}
\tanh\left ( \frac{J}{k_B \, T_c}\right ) = e^{ -   \frac{2\, \tilde J}{k_B \, T_c} } \,,
\end{align}
where we have restored the dimension-full quantities ($J, k_B$) for convenience of illustration.
Although this is not a spin glass system as it has no frustration, nevertheless it will be used to demonstrate that our method captures the $T_c$ even for a disordered system.

\section{Method}
\label{method}
Let us consider the Hamiltonian of a classical spin system defined on a square lattice.
We divide the lattice into two regions, $A$ and $B$,
with the spin configurations within each subsystem denoted by
$i_A$ and $i_B$ respectively. The probability of the state $i_A$ occurring in subregion $A$
is $p_{ i_ A }= \sum \limits_{i_B} p_{i_A ,i_B}$, where $  p_{i_A ,i_B}
=\frac{ e^{ - \beta E(i_A ,i_B )  } }{ Z [T]}  $
is the probability of existence of any arbitrary state of the entire system, obtained from the Boltzmann distribution.
We have denoted the energy associated with the states $i_A$ and $i_B$ by $E(i_A , i_B )$, and the partition function for $ A \cup B$ by $Z[T] 
=  \sum \limits_{i_A ,i_B} e^{ - \beta E(i_A ,i_B )} $, where $\beta^{-1} = {k_B\,T}$. We will use the units where $k_B$ is set to $1$.

In the partitioned system, the second R\'enyi entropy for subregion $A$
is defined by \cite{melko2010}:
\begin{align}
\label{s2}
S_2 (A)&=
- \ln   \left ( \sum_{ i_A}  p_{ i_ A }^2   \right )\nonumber \\
& =-\ln \left ( \sum_{ i_A}
\frac{  \sum \limits_{ i_B}  e^{ - \beta E(i_A ,i_B )}
 \,  \sum \limits_{ j_B}  e^{ - \beta E(i_A ,j_B )} ) }
{ Z^2 [T]}   \right ) \nonumber \\
& = -\ln \left ( Z[A,2,T] \right ) + 2 \ln \left ( Z[T] \right ) ,
\end{align}
where $Z [A, 2, T ] =  \sum \limits_{i_A,  i_B, j_B}  e^{ - \beta \lbrace E(i_A ,i_B ) + E(i_A ,j_B ) \rbrace}$
can be interpreted as the partition function of a new ``replicated'' system, such that the spins in subregion $A$ are restricted to be the same
in both the replicas, while the spins in subregion $B$ are unconstrained for the two copies. Due to the constraint for the spins in the replicas of the subregion $A$, there the spins in the bulk behave as if their effective temperature is $T/2$ for local interactions. Using the second R\'enyi entropies, the R{\'e}nyi mutual information (RMI) is defined as the symmetric function:
\begin{align}
\label{rmi}
I_2 (A, B) &= S_2 (A) + S_2 (B) - S_2 (A \cup B )\nonumber\\
& = -\ln \left (
\frac{  Z[ A,2, T] \,  Z[B, 2, T]
}
{  Z^2 [T] \,  Z[T/2] }
\right) \,.
\end{align}
The data is obtained by thermodynamic integration and imposing periodic boundary conditions on the lattice.

The RMI scales as \cite{Singh}:
\begin{align}
\frac{I_2(A,B)} {L} = a_2 (\beta) + \frac{d_2(\beta)}{L}  +\mathcal{O} \left(L^{-2}\right ),
\end{align}
where the term $d_2 (\beta)$ is related to the symmetry breaking of the lattice.
When the symmetry breaking causes $d_2 (\beta)$ to change sign, it passes through zero and the function $\frac{I_2(A,B)} {L}$ is then independent of system size up to order $\mathcal{O} \left(L^{-2}\right )$. One can show that $d_2(\beta) < 0$ for $T_c < T < 2 \, T_c$, and positive elsewhere. This leads to the crossings in $\frac{I_2 \left (L/2, L/2 \right )}{L}$ at the temperatures $T_c$ and $2 \,T_c$ for different system sizes. 
Thus the RMI as a function of temperature reveals a continuous phase transition at critical temperature $T_c$.
This method has been successfully applied for detecting finite temperature phase transitions with great accuracy in a variety of classical systems \cite{Singh,stephen2013,WL,ipsita-roger}.
\begin{figure}[ht]
\includegraphics[width=0.8\linewidth]{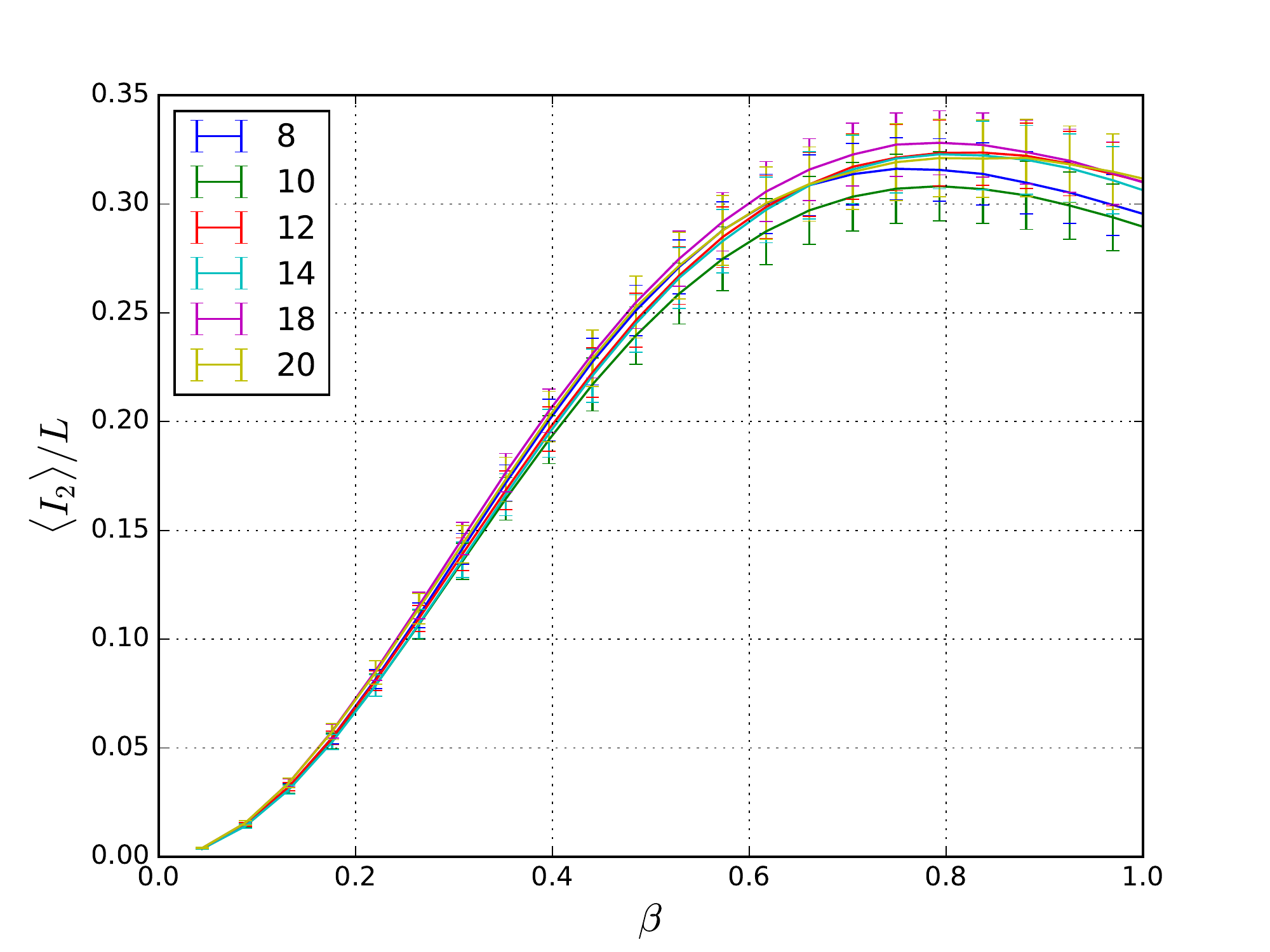}
\caption{\label{eafig}RMI for Edwards-Anderson model with $J_{ij}$ drawn from the continuous Gaussian bond distribution of Eq.~(\ref{prob1}) and averaged over $250$ disorder realizations.}
\end{figure}

\begin{figure}[ht]
\includegraphics[width=0.95\linewidth]{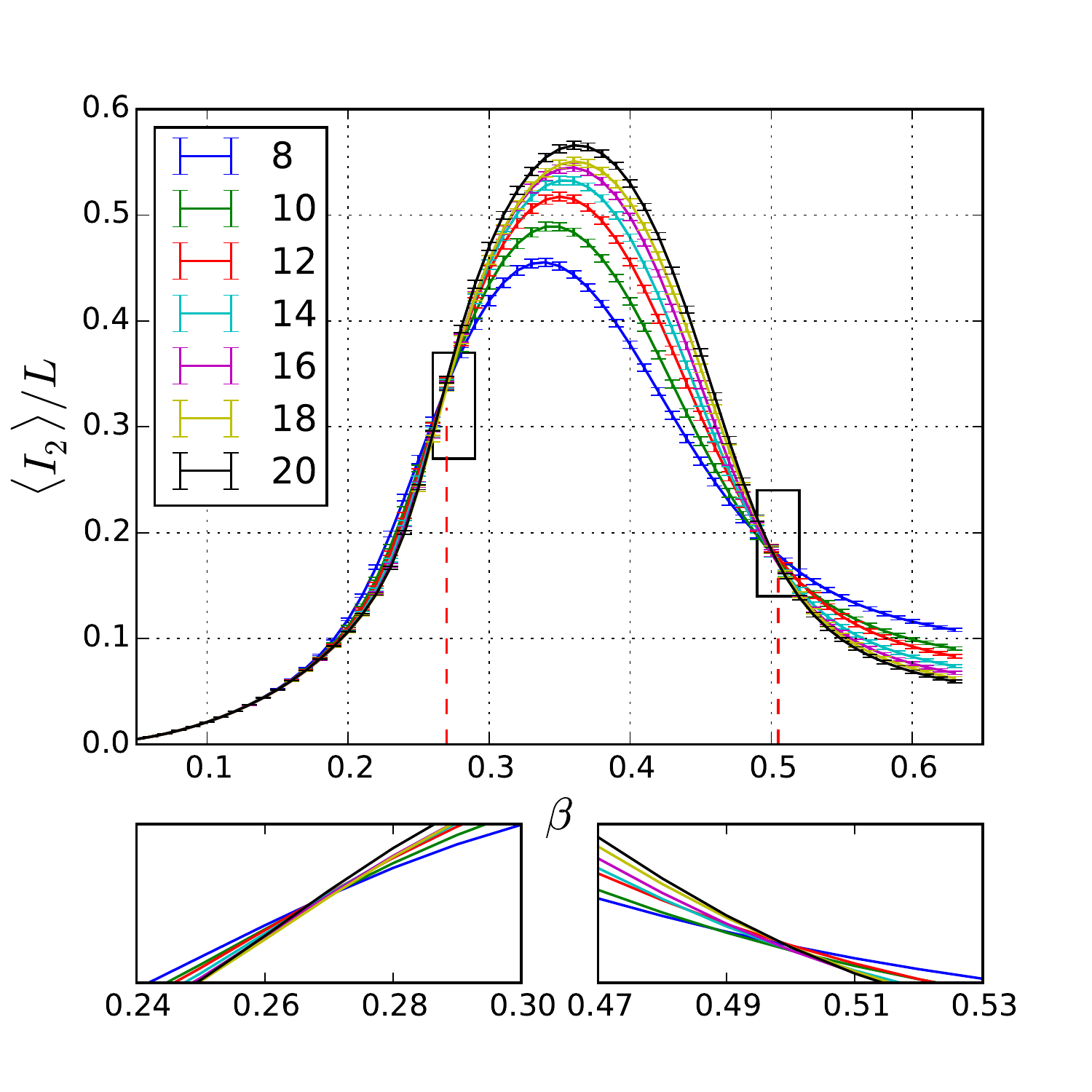}
\caption{
RMI with $p = 0.04$ for the distribution given in Eq.~\eqref{randomising}, averaged over $150$ disorder realizations. The crossings occur in the ranges $\beta \in [ 0.25, 0.28] $ and $\beta \in [0.49, 0.51]$.}
\label{fig:i3}
\end{figure}
\begin{figure*}[htb]
\subfigure[]{
\includegraphics[width=0.3\linewidth]{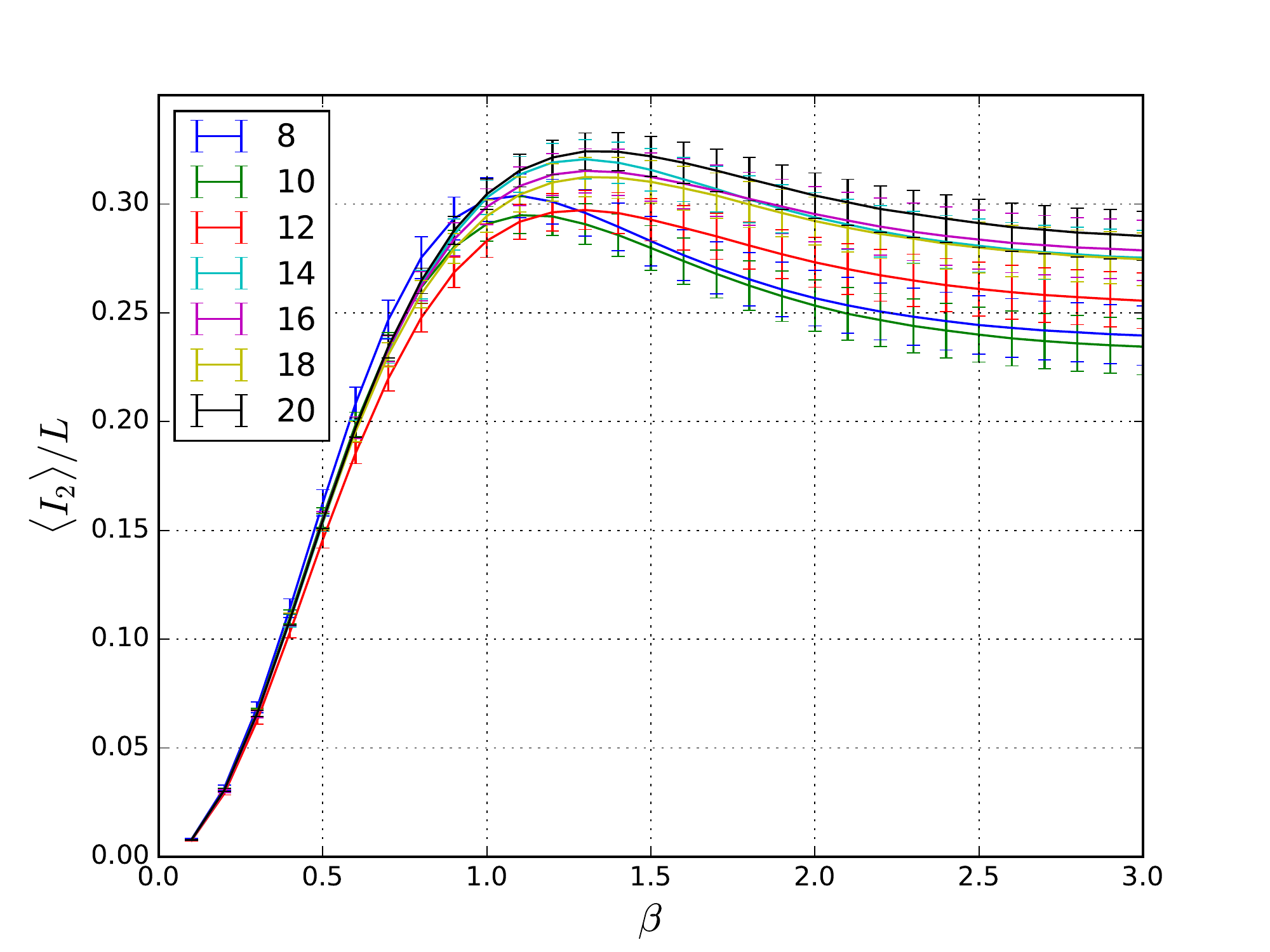}}
\subfigure[]{
\includegraphics[width=0.3\linewidth]{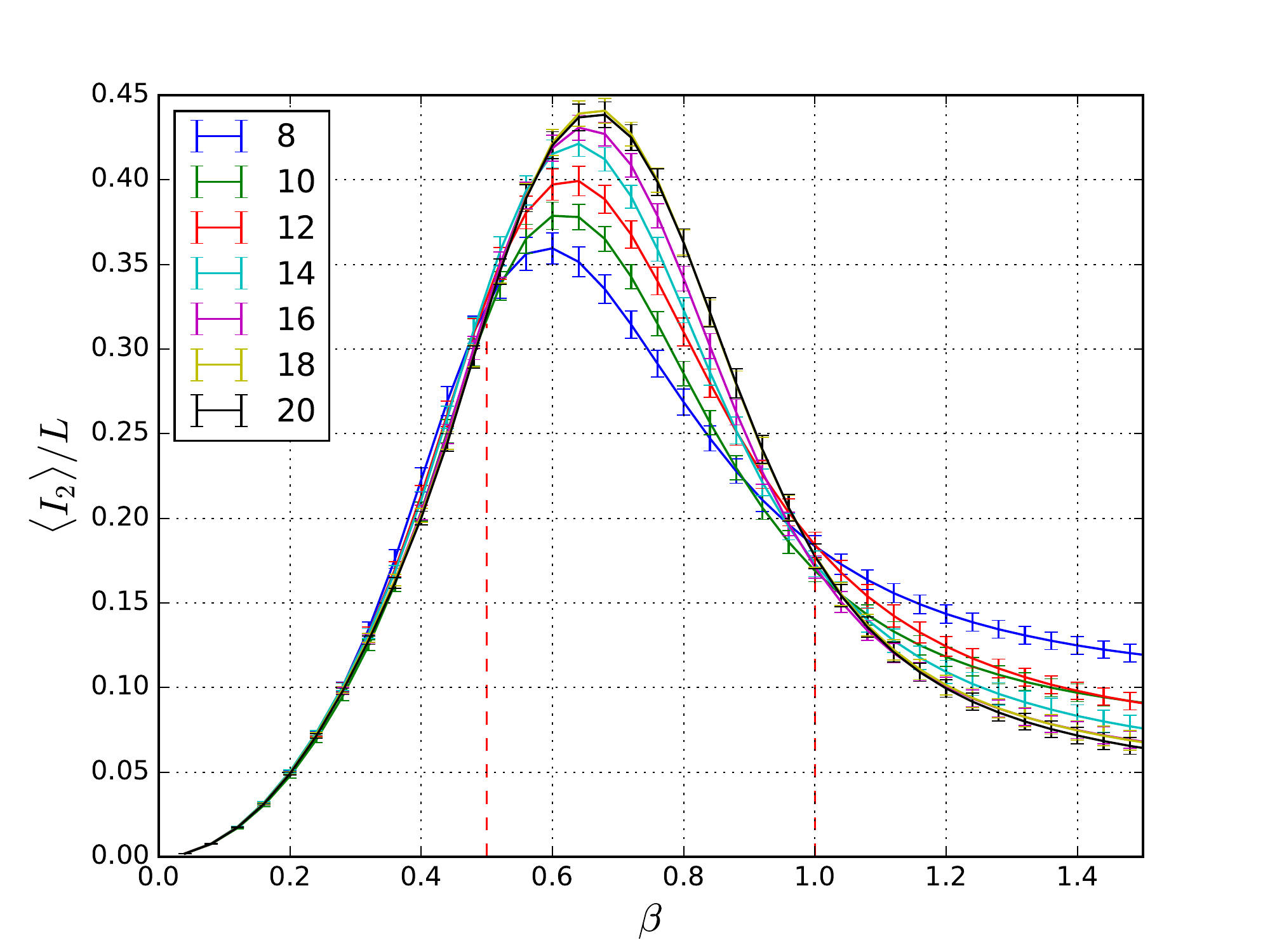}}
\subfigure[]{
\includegraphics[width=0.3\linewidth]{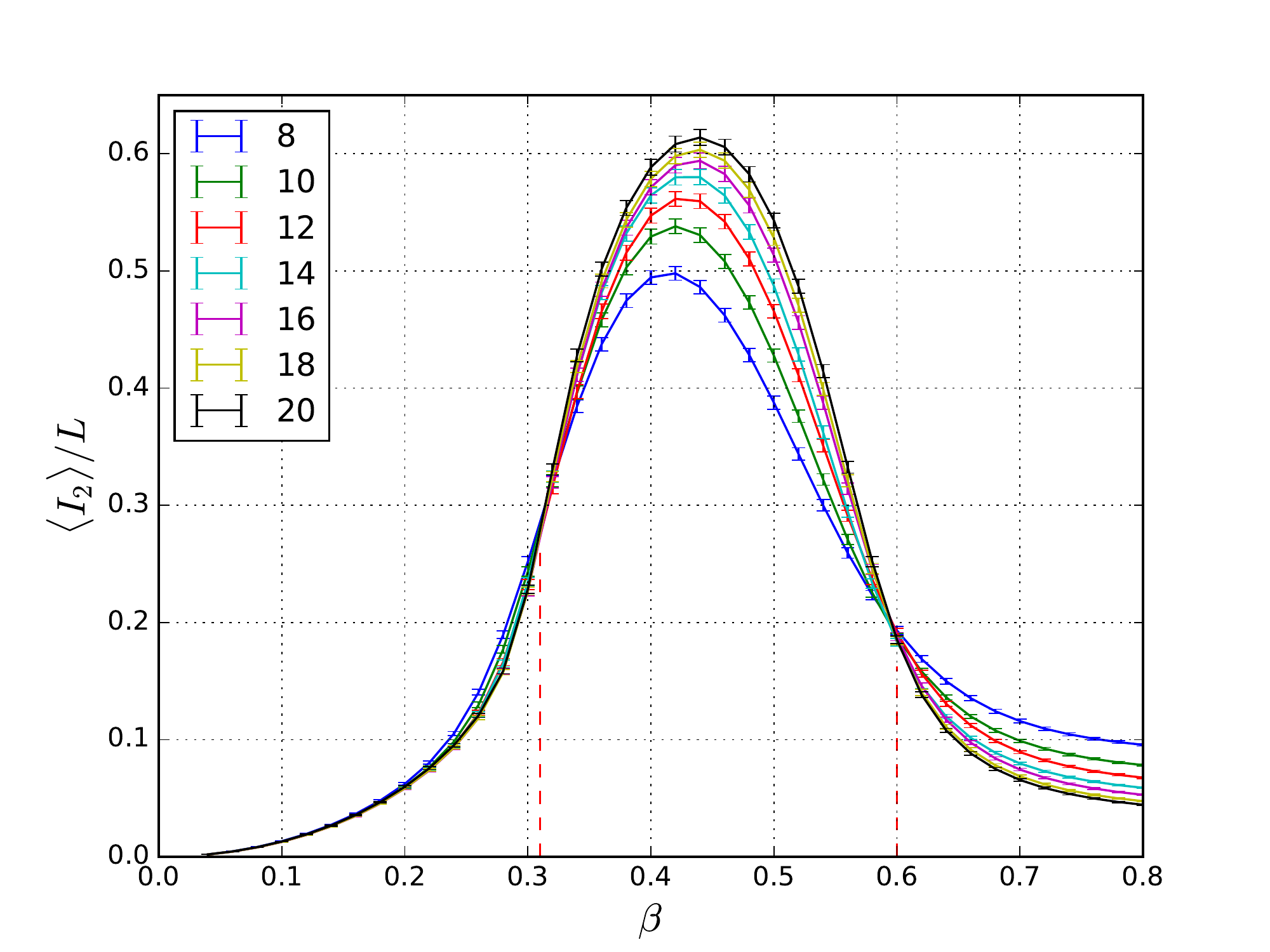}}
\caption{RMI for three different scenarios from Eq.~\eqref{diffcouplings}, averaged over $100$ disorder realizations. (a) For $\tilde J = 0$ and $p = 0.4$, no crossing is seen implying a zero-temperature phase transition.  (b) For $\tilde J = 0$ and $p = 0.6$, crossings occur at $\beta \approx 0.5 \pm 0.04$ and $ \beta  \approx 1.0 \pm 0.04$. (c) For $\tilde J = 0.5$ and $p = 0.5$ (self-dual point), crossings occur at $\beta  \approx 0.31 \pm 0.005$ and $ \beta  \approx  0.609 \pm 0.005$.}
\label{fig:model2}
\end{figure*}

In our numerical simulations, we compute the RMI using clasical Monte Carlo algorithm coupled with the transfer-matrix ratio trick \cite{gelman1998,tommaso,graph-theory}, from the expression
\begin{align}
\label{ratio}
 \frac{  Z [A,2, T] }   {  Z^2 [T] }
= \prod \limits_{i=0}^{N-1}  \frac{  Z [A_{ i+1 } ,2, T] }   {   Z [A_i ,2, T] } ,
Z [A_0 ,2, T]   = Z^2 [T] \,,
\end{align}
where $A_i$ denotes a series of $N$ blocks of increasing size, the consecutive blocks differing by a
one-dimensional strip of spins running parallel to the boundary separating $A$ and $B$. While $ A_0 $ labels the empty region, $A_N = A$. The details of the algorithm can be found in Ref.~\onlinecite{stephan2014}.  In addition to implementing this procedure, we supplement it by parallel tempering to ensure that the states used in the estimation of the ratios of the partition functions, $\Big \lbrace \frac{  Z [A_{ i+1 } ,2, T] }   {   Z [A_i ,2, T] }  \Big \rbrace $, are efficiently sampled. Finally, we find $\langle I_2 \rangle$ by averaging over a reasonable number of disorder realizations.

\section{Results}
\label{results}

In this section, we will demonstrate the behavior of the RMI curves for the different models described in Sec.~\ref{model}. The error bars for averaging over disorder realizations are indicated in each plot.


For the Edwards-Anderson model, we consider only the continuous bond distribution class given by Eq.~(\ref{prob1}). Consistent with the literature \cite{young-review}, the RMI curves in Fig.~\ref{eafig} do not show any single crossing point, indicating the absence of any finite temperature transition to the glassy phase.
The RCF model in Eq.~(\ref{ferro-model}) was predicted to have a nonzero $T_c$ \cite{ferro1,ferro2,ferro3}, but the geometry of the problem prevented certain Monte Carlo optimizations. We have used a cluster update algorithm for the Monte Carlo updates for interactions involving nearest neighbors. It is nontrivial to design such an algorithm for next nearest neighbor interactions, and hence it is beyond the scope of the current work. Without such cluster updates, the efficiency of the Monte Carlo simulations is order of magnitude less, and with the computation power available to us, this did not lead to convergent results.

For the random-bond Ising model with the distribution given in Eq.~\eqref{randomising}, we can clearly identify the $T_c$ associated with the presence of the ferromagnetic-paramagnetic phase transition for a representative case with $p=0.04$, as shown in Fig.~\ref{fig:i3}. The $\beta_c$ is found to be in the range $[0.49, 0.51]$, consistent with the literature \cite{aoki}.

Lastly, we show the results for the random-bond Ising model of Eq.~(\ref{diffcouplings}) only with ferromagnetic couplings in Fig.~\ref{fig:model2}. Again, the presence or absence of a nonzero $T_c$ is consistent with the literature \cite{jayaprakash,zobin,levy,fisch,derrida}. For the $p=1/2$ self-dual case, an analytic expression for the critical point exists (see Eq.~(\ref{analytic})). Solving Eq.~(\ref{analytic}) for $J=1$ and $\tilde J =0.5$, we obtain $\beta_c = 0.609 \pm 0.005$ which matches well with our numerical estimate of $\beta_c = 0.61$. These values are within error bars and more accuracy is achievable with smaller grid and more disorder realizations, which is limited by computational power available to us.

\section{Discussion and concluding remarks}
\label{conclude}

We have shown that the detection of classical phase transition points by the RMI method works seamlessly even for disordered systems. It was not at all clear {\it a priori} that one would be able to see the precise $I_2/L$ crossings at $T_c$ and $2\, T_c$, given the fact that the curves have finite error bars due to disorder averaging. Nonetheless, the method has successfully identified the $T_c$'s predicted in the literature for all the models considered, except the RCF.

Given the success of the RMI techniques to identify the phase transitions in these disordered models, in future works, one can apply an extension of this technique \cite{stephan2014,ipsita-roger} to extract the central charges of the CFTs associated with the corresponding critical points, which in many cases are either unknown or do not have an analytic expression.
The universality class can be identified if we can extract the central charge $c$ by using geometric mutual information.

\section{Acknowledgments} 
We thank Roger Melko and Stephen Inglis for collaboration in the initial stages of the project and valuable comments on the manuscript. We are also grateful to L. E. Hayward Sierens for useful discussions.

%
 

\begin{thebibliography}{34}%
\makeatletter
\providecommand \@ifxundefined [1]{%
 \@ifx{#1\undefined}
}%
\providecommand \@ifnum [1]{%
 \ifnum #1\expandafter \@firstoftwo
 \else \expandafter \@secondoftwo
 \fi
}%
\providecommand \@ifx [1]{%
 \ifx #1\expandafter \@firstoftwo
 \else \expandafter \@secondoftwo
 \fi
}%
\providecommand \natexlab [1]{#1}%
\providecommand \enquote  [1]{``#1''}%
\providecommand \bibnamefont  [1]{#1}%
\providecommand \bibfnamefont [1]{#1}%
\providecommand \citenamefont [1]{#1}%
\providecommand \href@noop [0]{\@secondoftwo}%
\providecommand \href [0]{\begingroup \@sanitize@url \@href}%
\providecommand \@href[1]{\@@startlink{#1}\@@href}%
\providecommand \@@href[1]{\endgroup#1\@@endlink}%
\providecommand \@sanitize@url [0]{\catcode `\\12\catcode `\$12\catcode
  `\&12\catcode `\#12\catcode `\^12\catcode `\_12\catcode `\%12\relax}%
\providecommand \@@startlink[1]{}%
\providecommand \@@endlink[0]{}%
\providecommand \url  [0]{\begingroup\@sanitize@url \@url }%
\providecommand \@url [1]{\endgroup\@href {#1}{\urlprefix }}%
\providecommand \urlprefix  [0]{URL }%
\providecommand \Eprint [0]{\href }%
\providecommand \doibase [0]{http://dx.doi.org/}%
\providecommand \selectlanguage [0]{\@gobble}%
\providecommand \bibinfo  [0]{\@secondoftwo}%
\providecommand \bibfield  [0]{\@secondoftwo}%
\providecommand \translation [1]{[#1]}%
\providecommand \BibitemOpen [0]{}%
\providecommand \bibitemStop [0]{}%
\providecommand \bibitemNoStop [0]{.\EOS\space}%
\providecommand \EOS [0]{\spacefactor3000\relax}%
\providecommand \BibitemShut  [1]{\csname bibitem#1\endcsname}%
\let\auto@bib@innerbib\@empty
\bibitem [{\citenamefont {Binder}\ and\ \citenamefont
  {Young}(1986)}]{young-review}%
  \BibitemOpen
  \bibfield  {author} {\bibinfo {author} {\bibfnamefont {K.}~\bibnamefont
  {Binder}}\ and\ \bibinfo {author} {\bibfnamefont {A.~P.}\ \bibnamefont
  {Young}},\ }\href {\doibase 10.1103/RevModPhys.58.801} {\bibfield  {journal}
  {\bibinfo  {journal} {Rev. Mod. Phys.}\ }\textbf {\bibinfo {volume} {58}},\
  \bibinfo {pages} {801} (\bibinfo {year} {1986})}\BibitemShut {NoStop}%
\bibitem [{\citenamefont {Nishimori}(1981)}]{nishimori}%
  \BibitemOpen
  \bibfield  {author} {\bibinfo {author} {\bibfnamefont {H.}~\bibnamefont
  {Nishimori}},\ }\href {\doibase 10.1143/PTP.66.1169} {\bibfield  {journal}
  {\bibinfo  {journal} {Progress of Theoretical Physics}\ }\textbf {\bibinfo
  {volume} {66}},\ \bibinfo {pages} {1169} (\bibinfo {year}
  {1981})}\BibitemShut {NoStop}%
\bibitem [{\citenamefont {Cho}\ and\ \citenamefont
  {Fisher}(1997)}]{matthew-fisher}%
  \BibitemOpen
  \bibfield  {author} {\bibinfo {author} {\bibfnamefont {S.}~\bibnamefont
  {Cho}}\ and\ \bibinfo {author} {\bibfnamefont {M.~P.~A.}\ \bibnamefont
  {Fisher}},\ }\href {\doibase 10.1103/PhysRevB.55.1025} {\bibfield  {journal}
  {\bibinfo  {journal} {Phys. Rev. B}\ }\textbf {\bibinfo {volume} {55}},\
  \bibinfo {pages} {1025} (\bibinfo {year} {1997})}\BibitemShut {NoStop}%
\bibitem [{\citenamefont {Fisch}(1976)}]{fisch}%
  \BibitemOpen
  \bibfield  {author} {\bibinfo {author} {\bibfnamefont {R.}~\bibnamefont
  {Fisch}},\ }\href {\doibase 10.1007/BF01014302} {\bibfield  {journal}
  {\bibinfo  {journal} {J Stat Phys}\ }\textbf {\bibinfo {volume} {18}},\
  \bibinfo {pages} {111} (\bibinfo {year} {1976})}\BibitemShut {NoStop}%
\bibitem [{\citenamefont {Jayaprakash}\ \emph {et~al.}(1978)\citenamefont
  {Jayaprakash}, \citenamefont {Riedel},\ and\ \citenamefont
  {Wortis}}]{jayaprakash}%
  \BibitemOpen
  \bibfield  {author} {\bibinfo {author} {\bibfnamefont {C.}~\bibnamefont
  {Jayaprakash}}, \bibinfo {author} {\bibfnamefont {E.~K.}\ \bibnamefont
  {Riedel}}, \ and\ \bibinfo {author} {\bibfnamefont {M.}~\bibnamefont
  {Wortis}},\ }\href {\doibase 10.1103/PhysRevB.18.2244} {\bibfield  {journal}
  {\bibinfo  {journal} {Phys. Rev. B}\ }\textbf {\bibinfo {volume} {18}},\
  \bibinfo {pages} {2244} (\bibinfo {year} {1978})}\BibitemShut {NoStop}%
\bibitem [{\citenamefont {Zobin}(1978)}]{zobin}%
  \BibitemOpen
  \bibfield  {author} {\bibinfo {author} {\bibfnamefont {D.}~\bibnamefont
  {Zobin}},\ }\href {\doibase 10.1103/PhysRevB.18.2387} {\bibfield  {journal}
  {\bibinfo  {journal} {Phys. Rev. B}\ }\textbf {\bibinfo {volume} {18}},\
  \bibinfo {pages} {2387} (\bibinfo {year} {1978})}\BibitemShut {NoStop}%
\bibitem [{\citenamefont {Tsallis}\ and\ \citenamefont {Levy}(1980)}]{levy}%
  \BibitemOpen
  \bibfield  {author} {\bibinfo {author} {\bibfnamefont {C.}~\bibnamefont
  {Tsallis}}\ and\ \bibinfo {author} {\bibfnamefont {S.~V.~F.}\ \bibnamefont
  {Levy}},\ }\href {http://stacks.iop.org/0022-3719/13/i=4/a=007} {\bibfield
  {journal} {\bibinfo  {journal} {Journal of Physics C: Solid State Physics}\
  }\textbf {\bibinfo {volume} {13}},\ \bibinfo {pages} {465} (\bibinfo {year}
  {1980})}\BibitemShut {NoStop}%
\bibitem [{\citenamefont {Melko}\ \emph {et~al.}(2010)\citenamefont {Melko},
  \citenamefont {Kallin},\ and\ \citenamefont {Hastings}}]{melko2010}%
  \BibitemOpen
  \bibfield  {author} {\bibinfo {author} {\bibfnamefont {R.~G.}\ \bibnamefont
  {Melko}}, \bibinfo {author} {\bibfnamefont {A.~B.}\ \bibnamefont {Kallin}}, \
  and\ \bibinfo {author} {\bibfnamefont {M.~B.}\ \bibnamefont {Hastings}},\
  }\href {\doibase 10.1103/PhysRevB.82.100409} {\bibfield  {journal} {\bibinfo
  {journal} {Phys. Rev. B}\ }\textbf {\bibinfo {volume} {82}},\ \bibinfo
  {pages} {100409} (\bibinfo {year} {2010})}\BibitemShut {NoStop}%
\bibitem [{\citenamefont {Singh}\ \emph {et~al.}(2011)\citenamefont {Singh},
  \citenamefont {Hastings}, \citenamefont {Kallin},\ and\ \citenamefont
  {Melko}}]{Singh}%
  \BibitemOpen
  \bibfield  {author} {\bibinfo {author} {\bibfnamefont {R.~R.~P.}\
  \bibnamefont {Singh}}, \bibinfo {author} {\bibfnamefont {M.~B.}\ \bibnamefont
  {Hastings}}, \bibinfo {author} {\bibfnamefont {A.~B.}\ \bibnamefont
  {Kallin}}, \ and\ \bibinfo {author} {\bibfnamefont {R.~G.}\ \bibnamefont
  {Melko}},\ }\href {\doibase 10.1103/PhysRevLett.106.135701} {\bibfield
  {journal} {\bibinfo  {journal} {Phys. Rev. Lett.}\ }\textbf {\bibinfo
  {volume} {106}},\ \bibinfo {pages} {135701} (\bibinfo {year}
  {2011})}\BibitemShut {NoStop}%
\bibitem [{\citenamefont {Inglis}\ and\ \citenamefont {Melko}(2013)}]{WL}%
  \BibitemOpen
  \bibfield  {author} {\bibinfo {author} {\bibfnamefont {S.}~\bibnamefont
  {Inglis}}\ and\ \bibinfo {author} {\bibfnamefont {R.~G.}\ \bibnamefont
  {Melko}},\ }\href {\doibase 10.1103/PhysRevE.87.013306} {\bibfield  {journal}
  {\bibinfo  {journal} {Phys. Rev. E}\ }\textbf {\bibinfo {volume} {87}},\
  \bibinfo {pages} {013306} (\bibinfo {year} {2013})}\BibitemShut {NoStop}%
\bibitem [{\citenamefont {Iaconis}\ \emph {et~al.}(2013)\citenamefont
  {Iaconis}, \citenamefont {Inglis}, \citenamefont {Kallin},\ and\
  \citenamefont {Melko}}]{stephen2013}%
  \BibitemOpen
  \bibfield  {author} {\bibinfo {author} {\bibfnamefont {J.}~\bibnamefont
  {Iaconis}}, \bibinfo {author} {\bibfnamefont {S.}~\bibnamefont {Inglis}},
  \bibinfo {author} {\bibfnamefont {A.~B.}\ \bibnamefont {Kallin}}, \ and\
  \bibinfo {author} {\bibfnamefont {R.~G.}\ \bibnamefont {Melko}},\ }\href
  {\doibase 10.1103/PhysRevB.87.195134} {\bibfield  {journal} {\bibinfo
  {journal} {Phys. Rev. B}\ }\textbf {\bibinfo {volume} {87}},\ \bibinfo
  {pages} {195134} (\bibinfo {year} {2013})}\BibitemShut {NoStop}%
\bibitem [{\citenamefont {St\'ephan}\ \emph {et~al.}(2014)\citenamefont
  {St\'ephan}, \citenamefont {Inglis}, \citenamefont {Fendley},\ and\
  \citenamefont {Melko}}]{stephan2014}%
  \BibitemOpen
  \bibfield  {author} {\bibinfo {author} {\bibfnamefont {J.-M.}\ \bibnamefont
  {St\'ephan}}, \bibinfo {author} {\bibfnamefont {S.}~\bibnamefont {Inglis}},
  \bibinfo {author} {\bibfnamefont {P.}~\bibnamefont {Fendley}}, \ and\
  \bibinfo {author} {\bibfnamefont {R.~G.}\ \bibnamefont {Melko}},\ }\href
  {\doibase 10.1103/PhysRevLett.112.127204} {\bibfield  {journal} {\bibinfo
  {journal} {Phys. Rev. Lett.}\ }\textbf {\bibinfo {volume} {112}},\ \bibinfo
  {pages} {127204} (\bibinfo {year} {2014})}\BibitemShut {NoStop}%
\bibitem [{\citenamefont {Wilms}\ \emph {et~al.}(2011)\citenamefont {Wilms},
  \citenamefont {Troyer},\ and\ \citenamefont {Verstraete}}]{troyer}%
  \BibitemOpen
  \bibfield  {author} {\bibinfo {author} {\bibfnamefont {J.}~\bibnamefont
  {Wilms}}, \bibinfo {author} {\bibfnamefont {M.}~\bibnamefont {Troyer}}, \
  and\ \bibinfo {author} {\bibfnamefont {F.}~\bibnamefont {Verstraete}},\
  }\href {http://stacks.iop.org/1742-5468/2011/i=10/a=P10011} {\bibfield
  {journal} {\bibinfo  {journal} {Journal of Statistical Mechanics: Theory and
  Experiment}\ }\textbf {\bibinfo {volume} {2011}},\ \bibinfo {pages} {P10011}
  (\bibinfo {year} {2011})}\BibitemShut {NoStop}%
\bibitem [{\citenamefont {Wilms}\ \emph {et~al.}(2012)\citenamefont {Wilms},
  \citenamefont {Vidal}, \citenamefont {Verstraete},\ and\ \citenamefont
  {Dusuel}}]{vidal}%
  \BibitemOpen
  \bibfield  {author} {\bibinfo {author} {\bibfnamefont {J.}~\bibnamefont
  {Wilms}}, \bibinfo {author} {\bibfnamefont {J.}~\bibnamefont {Vidal}},
  \bibinfo {author} {\bibfnamefont {F.}~\bibnamefont {Verstraete}}, \ and\
  \bibinfo {author} {\bibfnamefont {S.}~\bibnamefont {Dusuel}},\ }\href
  {http://stacks.iop.org/1742-5468/2012/i=01/a=P01023} {\bibfield  {journal}
  {\bibinfo  {journal} {Journal of Statistical Mechanics: Theory and
  Experiment}\ }\textbf {\bibinfo {volume} {2012}},\ \bibinfo {pages} {P01023}
  (\bibinfo {year} {2012})}\BibitemShut {NoStop}%
\bibitem [{\citenamefont {Alba}(2013)}]{Alba1}%
  \BibitemOpen
  \bibfield  {author} {\bibinfo {author} {\bibfnamefont {V.}~\bibnamefont
  {Alba}},\ }\href@noop {} {\bibfield  {journal} {\bibinfo  {journal} {Journal
  of Statistical Mechanics: Theory and Experiment}\ }\textbf {\bibinfo {volume}
  {2013}},\ \bibinfo {pages} {P05013} (\bibinfo {year} {2013})}\BibitemShut
  {NoStop}%
\bibitem [{\citenamefont {Chung}\ \emph {et~al.}(2014)\citenamefont {Chung},
  \citenamefont {Alba}, \citenamefont {Bonnes}, \citenamefont {Chen},\ and\
  \citenamefont {L\"auchli}}]{Alba2}%
  \BibitemOpen
  \bibfield  {author} {\bibinfo {author} {\bibfnamefont {C.-M.}\ \bibnamefont
  {Chung}}, \bibinfo {author} {\bibfnamefont {V.}~\bibnamefont {Alba}},
  \bibinfo {author} {\bibfnamefont {L.}~\bibnamefont {Bonnes}}, \bibinfo
  {author} {\bibfnamefont {P.}~\bibnamefont {Chen}}, \ and\ \bibinfo {author}
  {\bibfnamefont {A.~M.}\ \bibnamefont {L\"auchli}},\ }\href {\doibase
  10.1103/PhysRevB.90.064401} {\bibfield  {journal} {\bibinfo  {journal} {Phys.
  Rev. B}\ }\textbf {\bibinfo {volume} {90}},\ \bibinfo {pages} {064401}
  (\bibinfo {year} {2014})}\BibitemShut {NoStop}%
\bibitem [{\citenamefont {Alba}\ \emph {et~al.}(2016)\citenamefont {Alba},
  \citenamefont {Inglis},\ and\ \citenamefont {Pollet}}]{stephen2016}%
  \BibitemOpen
  \bibfield  {author} {\bibinfo {author} {\bibfnamefont {V.}~\bibnamefont
  {Alba}}, \bibinfo {author} {\bibfnamefont {S.}~\bibnamefont {Inglis}}, \ and\
  \bibinfo {author} {\bibfnamefont {L.}~\bibnamefont {Pollet}},\ }\href
  {\doibase 10.1103/PhysRevB.93.094404} {\bibfield  {journal} {\bibinfo
  {journal} {Phys. Rev. B}\ }\textbf {\bibinfo {volume} {93}},\ \bibinfo
  {pages} {094404} (\bibinfo {year} {2016})}\BibitemShut {NoStop}%
\bibitem [{\citenamefont {Helmes}\ \emph {et~al.}(2015)\citenamefont {Helmes},
  \citenamefont {St\'ephan},\ and\ \citenamefont {Trebst}}]{Johannes}%
  \BibitemOpen
  \bibfield  {author} {\bibinfo {author} {\bibfnamefont {J.}~\bibnamefont
  {Helmes}}, \bibinfo {author} {\bibfnamefont {J.-M.}\ \bibnamefont
  {St\'ephan}}, \ and\ \bibinfo {author} {\bibfnamefont {S.}~\bibnamefont
  {Trebst}},\ }\href {\doibase 10.1103/PhysRevB.92.125144} {\bibfield
  {journal} {\bibinfo  {journal} {Phys. Rev. B}\ }\textbf {\bibinfo {volume}
  {92}},\ \bibinfo {pages} {125144} (\bibinfo {year} {2015})}\BibitemShut
  {NoStop}%
\bibitem [{\citenamefont {Mandal}\ \emph {et~al.}(2016)\citenamefont {Mandal},
  \citenamefont {Inglis},\ and\ \citenamefont {Melko}}]{ipsita-roger}%
  \BibitemOpen
  \bibfield  {author} {\bibinfo {author} {\bibfnamefont {I.}~\bibnamefont
  {Mandal}}, \bibinfo {author} {\bibfnamefont {S.}~\bibnamefont {Inglis}}, \
  and\ \bibinfo {author} {\bibfnamefont {R.~G.}\ \bibnamefont {Melko}},\ }\href
  {http://stacks.iop.org/1742-5468/2016/i=7/a=073105} {\bibfield  {journal}
  {\bibinfo  {journal} {Journal of Statistical Mechanics: Theory and
  Experiment}\ }\textbf {\bibinfo {volume} {2016}},\ \bibinfo {pages} {073105}
  (\bibinfo {year} {2016})}\BibitemShut {NoStop}%
\bibitem [{\citenamefont {Shannon}(2001)}]{shannon}%
  \BibitemOpen
  \bibfield  {author} {\bibinfo {author} {\bibfnamefont {C.~E.}\ \bibnamefont
  {Shannon}},\ }\href@noop {} {\bibfield  {journal} {\bibinfo  {journal} {ACM
  SIGMOBILE Mobile Computing and Communications Review}\ }\textbf {\bibinfo
  {volume} {5}},\ \bibinfo {pages} {3} (\bibinfo {year} {2001})}\BibitemShut
  {NoStop}%
\bibitem [{\citenamefont {Calabrese}\ \emph {et~al.}(2009)\citenamefont
  {Calabrese}, \citenamefont {Cardy},\ and\ \citenamefont {Doyon}}]{cardy}%
  \BibitemOpen
  \bibfield  {author} {\bibinfo {author} {\bibfnamefont {P.}~\bibnamefont
  {Calabrese}}, \bibinfo {author} {\bibfnamefont {J.}~\bibnamefont {Cardy}}, \
  and\ \bibinfo {author} {\bibfnamefont {B.}~\bibnamefont {Doyon}},\ }\href
  {http://stacks.iop.org/1751-8121/42/i=50/a=500301} {\bibfield  {journal}
  {\bibinfo  {journal} {Journal of Physics A: Mathematical and Theoretical}\
  }\textbf {\bibinfo {volume} {42}},\ \bibinfo {pages} {500301} (\bibinfo
  {year} {2009})}\BibitemShut {NoStop}%
\bibitem [{\citenamefont {Belavin}\ \emph {et~al.}(1984)\citenamefont
  {Belavin}, \citenamefont {Polyakov},\ and\ \citenamefont
  {Zamolodchikov}}]{belavin}%
  \BibitemOpen
  \bibfield  {author} {\bibinfo {author} {\bibfnamefont {A.}~\bibnamefont
  {Belavin}}, \bibinfo {author} {\bibfnamefont {A.}~\bibnamefont {Polyakov}}, \
  and\ \bibinfo {author} {\bibfnamefont {A.}~\bibnamefont {Zamolodchikov}},\
  }\href {\doibase http://dx.doi.org/10.1016/0550-3213(84)90052-X} {\bibfield
  {journal} {\bibinfo  {journal} {Nuclear Physics B}\ }\textbf {\bibinfo
  {volume} {241}},\ \bibinfo {pages} {333 } (\bibinfo {year}
  {1984})}\BibitemShut {NoStop}%
\bibitem [{\citenamefont {Friedan}\ \emph {et~al.}(1984)\citenamefont
  {Friedan}, \citenamefont {Qiu},\ and\ \citenamefont {Shenker}}]{friedan}%
  \BibitemOpen
  \bibfield  {author} {\bibinfo {author} {\bibfnamefont {D.}~\bibnamefont
  {Friedan}}, \bibinfo {author} {\bibfnamefont {Z.}~\bibnamefont {Qiu}}, \ and\
  \bibinfo {author} {\bibfnamefont {S.}~\bibnamefont {Shenker}},\ }\href
  {\doibase 10.1103/PhysRevLett.52.1575} {\bibfield  {journal} {\bibinfo
  {journal} {Phys. Rev. Lett.}\ }\textbf {\bibinfo {volume} {52}},\ \bibinfo
  {pages} {1575} (\bibinfo {year} {1984})}\BibitemShut {NoStop}%
\bibitem [{\citenamefont {Holzhey}\ \emph {et~al.}(1994)\citenamefont
  {Holzhey}, \citenamefont {Larsen},\ and\ \citenamefont {Wilczek}}]{wilczek}%
  \BibitemOpen
  \bibfield  {author} {\bibinfo {author} {\bibfnamefont {C.}~\bibnamefont
  {Holzhey}}, \bibinfo {author} {\bibfnamefont {F.}~\bibnamefont {Larsen}}, \
  and\ \bibinfo {author} {\bibfnamefont {F.}~\bibnamefont {Wilczek}},\ }\href
  {\doibase http://dx.doi.org/10.1016/0550-3213(94)90402-2} {\bibfield
  {journal} {\bibinfo  {journal} {Nuclear Physics B}\ }\textbf {\bibinfo
  {volume} {424}},\ \bibinfo {pages} {443 } (\bibinfo {year}
  {1994})}\BibitemShut {NoStop}%
\bibitem [{\citenamefont {Vidal}\ \emph {et~al.}(2003)\citenamefont {Vidal},
  \citenamefont {Latorre}, \citenamefont {Rico},\ and\ \citenamefont
  {Kitaev}}]{kitaev}%
  \BibitemOpen
  \bibfield  {author} {\bibinfo {author} {\bibfnamefont {G.}~\bibnamefont
  {Vidal}}, \bibinfo {author} {\bibfnamefont {J.~I.}\ \bibnamefont {Latorre}},
  \bibinfo {author} {\bibfnamefont {E.}~\bibnamefont {Rico}}, \ and\ \bibinfo
  {author} {\bibfnamefont {A.}~\bibnamefont {Kitaev}},\ }\href {\doibase
  10.1103/PhysRevLett.90.227902} {\bibfield  {journal} {\bibinfo  {journal}
  {Phys. Rev. Lett.}\ }\textbf {\bibinfo {volume} {90}},\ \bibinfo {pages}
  {227902} (\bibinfo {year} {2003})}\BibitemShut {NoStop}%
\bibitem [{\citenamefont {Edwards}\ and\ \citenamefont
  {Anderson}(1975)}]{edwards}%
  \BibitemOpen
  \bibfield  {author} {\bibinfo {author} {\bibfnamefont {S.~F.}\ \bibnamefont
  {Edwards}}\ and\ \bibinfo {author} {\bibfnamefont {P.~W.}\ \bibnamefont
  {Anderson}},\ }\href {http://stacks.iop.org/0305-4608/5/i=5/a=017} {\bibfield
   {journal} {\bibinfo  {journal} {Journal of Physics F: Metal Physics}\
  }\textbf {\bibinfo {volume} {5}},\ \bibinfo {pages} {965} (\bibinfo {year}
  {1975})}\BibitemShut {NoStop}%
\bibitem [{\citenamefont {{Kawashima}}\ and\ \citenamefont
  {{Aoki}}(1999)}]{aoki}%
  \BibitemOpen
  \bibfield  {author} {\bibinfo {author} {\bibfnamefont {N.}~\bibnamefont
  {{Kawashima}}}\ and\ \bibinfo {author} {\bibfnamefont {T.}~\bibnamefont
  {{Aoki}}},\ }\href@noop {} {\bibfield  {journal} {\bibinfo  {journal} {eprint
  arXiv:cond-mat/9911120}\ } (\bibinfo {year} {1999})},\ \Eprint
  {http://arxiv.org/abs/cond-mat/9911120} {cond-mat/9911120} \BibitemShut
  {NoStop}%
\bibitem [{\citenamefont {Lemke}\ and\ \citenamefont
  {Campbell}(1996)}]{ferro1}%
  \BibitemOpen
  \bibfield  {author} {\bibinfo {author} {\bibfnamefont {N.}~\bibnamefont
  {Lemke}}\ and\ \bibinfo {author} {\bibfnamefont {I.~A.}\ \bibnamefont
  {Campbell}},\ }\href {\doibase 10.1103/PhysRevLett.76.4616} {\bibfield
  {journal} {\bibinfo  {journal} {Phys. Rev. Lett.}\ }\textbf {\bibinfo
  {volume} {76}},\ \bibinfo {pages} {4616} (\bibinfo {year}
  {1996})}\BibitemShut {NoStop}%
\bibitem [{\citenamefont {Lemke}\ and\ \citenamefont
  {Campbell}(1999)}]{ferro2}%
  \BibitemOpen
  \bibfield  {author} {\bibinfo {author} {\bibfnamefont {N.}~\bibnamefont
  {Lemke}}\ and\ \bibinfo {author} {\bibfnamefont {I.~A.}\ \bibnamefont
  {Campbell}},\ }\href {http://stacks.iop.org/0305-4470/32/i=45/a=304}
  {\bibfield  {journal} {\bibinfo  {journal} {Journal of Physics A:
  Mathematical and General}\ }\textbf {\bibinfo {volume} {32}},\ \bibinfo
  {pages} {7851} (\bibinfo {year} {1999})}\BibitemShut {NoStop}%
\bibitem [{\citenamefont {Hartmann}\ and\ \citenamefont
  {Campbell}(2001)}]{ferro3}%
  \BibitemOpen
  \bibfield  {author} {\bibinfo {author} {\bibfnamefont {A.~K.}\ \bibnamefont
  {Hartmann}}\ and\ \bibinfo {author} {\bibfnamefont {I.~A.}\ \bibnamefont
  {Campbell}},\ }\href {\doibase 10.1103/PhysRevB.63.094423} {\bibfield
  {journal} {\bibinfo  {journal} {Phys. Rev. B}\ }\textbf {\bibinfo {volume}
  {63}},\ \bibinfo {pages} {094423} (\bibinfo {year} {2001})}\BibitemShut
  {NoStop}%
\bibitem [{\citenamefont {Derrida}\ \emph {et~al.}(1987)\citenamefont
  {Derrida}, \citenamefont {Southern},\ and\ \citenamefont
  {Stauffer}}]{derrida}%
  \BibitemOpen
  \bibfield  {author} {\bibinfo {author} {\bibfnamefont {R.~B.}\ \bibnamefont
  {Derrida}}, \bibinfo {author} {\bibfnamefont {B.}~\bibnamefont {Southern}}, \
  and\ \bibinfo {author} {\bibfnamefont {D.}~\bibnamefont {Stauffer}},\ }\href
  {\doibase 10.1051/jphys:01987004803033500} {\bibfield  {journal} {\bibinfo
  {journal} {J. Phys. France}\ }\textbf {\bibinfo {volume} {48}},\ \bibinfo
  {pages} {335} (\bibinfo {year} {1987})}\BibitemShut {NoStop}%
\bibitem [{\citenamefont {Gelman}\ and\ \citenamefont
  {Meng}(1998)}]{gelman1998}%
  \BibitemOpen
  \bibfield  {author} {\bibinfo {author} {\bibfnamefont {A.}~\bibnamefont
  {Gelman}}\ and\ \bibinfo {author} {\bibfnamefont {X.-L.}\ \bibnamefont
  {Meng}},\ }\href {\doibase 10.1214/ss/1028905934} {\bibfield  {journal}
  {\bibinfo  {journal} {Statist. Sci.}\ }\textbf {\bibinfo {volume} {13}},\
  \bibinfo {pages} {163} (\bibinfo {year} {1998})}\BibitemShut {NoStop}%
\bibitem [{\citenamefont {Humeniuk}\ and\ \citenamefont
  {Roscilde}(2012)}]{tommaso}%
  \BibitemOpen
  \bibfield  {author} {\bibinfo {author} {\bibfnamefont {S.}~\bibnamefont
  {Humeniuk}}\ and\ \bibinfo {author} {\bibfnamefont {T.}~\bibnamefont
  {Roscilde}},\ }\href {\doibase 10.1103/PhysRevB.86.235116} {\bibfield
  {journal} {\bibinfo  {journal} {Phys. Rev. B}\ }\textbf {\bibinfo {volume}
  {86}},\ \bibinfo {pages} {235116} (\bibinfo {year} {2012})}\BibitemShut
  {NoStop}%
\bibitem [{\citenamefont {{Molkaraie}}\ and\ \citenamefont
  {{Loeliger}}(2013)}]{graph-theory}%
  \BibitemOpen
  \bibfield  {author} {\bibinfo {author} {\bibfnamefont {M.}~\bibnamefont
  {{Molkaraie}}}\ and\ \bibinfo {author} {\bibfnamefont {H.-A.}\ \bibnamefont
  {{Loeliger}}},\ }\href@noop {} {\bibfield  {journal} {\bibinfo  {journal}
  {ArXiv e-prints}\ } (\bibinfo {year} {2013})},\ \Eprint
  {http://arxiv.org/abs/1307.3645} {arXiv:1307.3645 [cs.IT]} \BibitemShut
  {NoStop}%
\end{thebibliography}
\end{document}